\documentclass{llncs} 
\usepackage{graphicx} 
\usepackage{wrapfig}
\usepackage{latexsym}
\usepackage{url}
\usepackage{xspace}
\usepackage{amssymb}
\usepackage{amsmath}
\usepackage{ifthen}
\usepackage{color}
\usepackage{alltt}
\usepackage[table]{xcolor}

\begin{document}
\author{Florian Lonsing \and Uwe Egly}
\institute{Knowledge-Based Systems Group
  \\ Vienna University of Technology, Austria \\ \url{http://www.kr.tuwien.ac.at/}}
\title{Incrementally Computing Minimal Unsatisfiable Cores of QBFs via a
Clause Group Solver API\thanks{Supported by the Austrian Science Fund (FWF)
under grant S11409-N23. We would like to thank Aina Niemetz and Mathias
Preiner for helpful discussions. This article will appear in the proceedings
of the \emph{18th International Conference on Theory and Applications of
  Satisfiability Testing (SAT)}, LNCS, Springer, 2015.}}
\maketitle
\begin{abstract}
We consider the incremental computation of minimal unsatisfiable cores (MUCs)
of QBFs. To this end, we equipped our incremental QBF solver DepQBF with a
novel API to allow for incremental solving based on clause groups. A clause
group is a set of clauses which is incrementally added to or removed from a
previously solved QBF. Our implementation of the novel API is related to
incremental SAT solving based on selector variables and assumptions. However,
the API entirely hides selector variables and assumptions from the user, which
facilitates the integration of DepQBF in other tools. We present implementation details
and, for the first time, report on experiments related to the 
computation of MUCs of QBFs using DepQBF's novel clause group API.
\end{abstract}


\section{Introduction}

Let $\psi = \hat{Q}.\,\phi$ be a QBF in \emph{prenex CNF (PCNF)} where
$\hat{Q} = Q_1x_1 \ldots Q_nx_n$ with $Q_i \in \{\forall,\exists\}$ is the
prefix containing quantified propositional variables $x_i$ and $\phi$ is a
quantifier-free CNF. Given a PCNF $\psi = \hat{Q}.\,\phi$, an
\emph{unsatisfiable core (UC)} of $\psi$ is an unsatisfiable PCNF $\psi' =
\hat{Q}'.\,\phi'$ such that $\hat{Q}' \subseteq \hat{Q}$ and $\phi' \subseteq
\phi$. The prefix $\hat{Q}'$ is obtained from $\hat{Q}$ by deleting the
quantified variables which do not occur in $\phi'$. A \emph{minimal
unsatisfiable core (MUC)}\footnote{The terminology \emph{minimal unsatisfiable
subformula (MUS)} is equivalent to MUC.} of $\psi$ is an unsatisfiable core
$\psi' = \hat{Q}'.\,\phi'$ of $\psi$ where, for every $C \in \phi'$, the PCNF
$\hat{Q}'.\,(\phi' \setminus \{C\})$ is satisfiable.

\emph{Incremental solving} is crucial for the computation of MUCs in the
context of propositional logic (SAT),
e.g.~\cite{DBLP:conf/lpar/AsinNOR08,DBLP:journals/aicom/BelovLM12,DBLP:conf/sat/DershowitzHN06,DBLP:conf/ictai/GregoireMP08,DBLP:conf/ismvl/Silva10,DBLP:conf/sat/SilvaL11,DBLP:journals/jsat/NadelRS14}.
Modifications of a CNF by adding and deleting clauses in incremental solving
are typically implemented by \emph{selector variables} and
\emph{assumptions}~\cite{DBLP:conf/sat/AudemardLS13,DBLP:conf/sat/EenS03,DBLP:journals/entcs/EenS03,DBLP:conf/sat/LagniezB13,DBLP:conf/cp/LonsingE14,DBLP:conf/icms/LonsingE14,DBLP:conf/date/MarinMLB12,DBLP:journals/aicom/MillerMB15,DBLP:conf/sat/NadelRS14}. An
added clause $C$ is augmented by a fresh selector variable $s$ so that
actually $C \cup \{s\}$ is added. Via the solver API, the user assigns these
variables as assumptions under which the CNF is solved to control whether a
clause is effectively present \nolinebreak in \nolinebreak the \nolinebreak
CNF.

Different from the assumption-based approach, the SAT solver
zChaff\footnote{zChaff website (July 2015):
\url{https://www.princeton.edu/~chaff/zchaff.html}}~\cite{DBLP:conf/dac/MoskewiczMZZM01}
provides an API to modify the CNF by adding and removing \emph{groups} (sets)
of clauses. Clauses are associated with an integer ID of the group they belong
to.

In assumption-based incremental solving, clause groups may be emulated by
augmenting all clauses in a group by the same selector variable. The user must
specify the necessary assumptions via the API in all forthcoming solver
invocations to enable and disable the right groups. In contrast to that,
zChaff allows to delete groups by a single API function call.  In terms of
usability, we argue that incremental solving by a clause group API is less
error-prone, more accessible to inexperienced users, and facilitates the
integration of the \nolinebreak solver \nolinebreak in \nolinebreak other
tools.

We present a novel clause group API of our QBF solver DepQBF (version 4.0
or later)\footnote{DepQBF is free software:
\url{http://lonsing.github.io/depqbf/}} in the style of zChaff. Different from zChaff, we implemented
clause groups based on selector variables and assumptions to combine the
conceptual simplicity of zChaff's API with state of the art assumption-based
incremental solving.  As a novel feature of our API, the handling of selector
variables and assumptions is entirely carried out by the solver and is hidden
from the user. Our approach is applicable to any SAT or QBF solver
supporting assumptions.  Based on the novel clause group API of DepQBF, we
implemented a tool to compute MUCs of PCNFs, a problem which 
has not been considered so far. Results on benchmarks used in the QBF Gallery
2014 illustrate the applicability of the clause group API for MUC
computation of PCNFs.


\section{Implementing a Clause Group API}

DepQBF is a solver for PCNFs based on the QBF-specific variant of the DPLL
algorithm~\cite{DBLP:journals/jar/CadoliSGG02} with
learning~\cite{DBLP:journals/jair/GiunchigliaNT06,DBLP:conf/tableaux/Letz02,DBLP:conf/cp/ZhangM02}.
Since version~3.0~\cite{DBLP:conf/cp/LonsingE14,DBLP:conf/icms/LonsingE14},
DepQBF supports incremental QBF solving via an API to add and remove clauses
in a stack-based way (cf.~Fig.~3 in \cite{DBLP:conf/icms/LonsingE14}). This
API is suitable for solving incremental encodings where clauses added most
recently tend to be removed again in subsequent solver calls, like
reachability problems such as conformant
planning~\cite{DBLP:conf/aisc/EglyKLP14} or bounded model
checking~\cite{jsat:benedetti08,DBLP:journals/entcs/JussilaB07}.  The new
clause group API of DepQBF, however, allows to add and delete clauses
\emph{arbitrarily}, which is necessary for the incremental computation of MUCs
of PCNFs.  We first present our novel approach to keeping selector variables
invisible to the user, which is a unique feature of DepQBF. To this end, we
distinguish between selector variables and variables in the encoding.

Let $S = \langle \psi_1,\ldots,\psi_n\rangle$ be a sequence of PCNFs.  We
consider variables over which the PCNFs $\psi_i$ are defined as \emph{user
variables} because they are part of the problem encoding represented by
$S$. When solving $S$ incrementally, \emph{selector variables} used to augment
clauses in $\psi_i$ are not part of the original encoding.  Variables $v$ are
stored in an array $\mathit{VA}$ indexed by an integer ID $\mathit{id}(v)$ of
$v$ such that $\mathit{VA}[\mathit{id}(v)] = v$. User and selector variables
reside in separate sections of $\mathit{VA}$:
{
\begin{center}
\raisebox{0.335cm}{$\mathit{VA}$:\ } \begin{tabular}{|c|c|c|c|c|c|c|c|c|c|c|c|} 
\hline
0 & 1 & \multicolumn{3}{c|}{\ldots \hspace{1.5cm}\ldots} & $\mathit{us} - 1$
& \cellcolor{black} \phantom{x} & $\mathit{us}$ & $\mathit{us} + 1$ &
\multicolumn{2}{c|}{\ldots \hspace{1.5cm}\ldots} & $\mathit{vs} - 1$\\
\hline
\multicolumn{6}{c}{\raisebox{0.25cm}{$\underbrace{\hspace{4.25cm}}$}} & \multicolumn{1}{c}{} & \multicolumn{5}{c}{\raisebox{0.25cm}{$\underbrace{\hspace{5.35cm}}$}} \\[-0.15cm]
\multicolumn{6}{c}{\raisebox{0.0cm}{\emph{user variables}}} &
\multicolumn{1}{c}{} & \multicolumn{5}{c}{\raisebox{0.0cm}{\emph{selector variables}}} \\
\end{tabular}
\end{center}
}
The total size of $\mathit{VA}$ is $\mathit{vs}$. The user variable section
has size $\mathit{us}$.  The following invariants are maintained:
$\mathit{VA}[\mathit{id}(v)] = v$ where $\mathit{id}(v) < \mathit{us}$ if $v$
is a user variable and $\mathit{us} \leq \mathit{id}(v) < \mathit{vs}$ if $v$
is a selector variable. If a new user variable $v$ with $\mathit{id}(v) \geq
\mathit{us}$ is added via the solver API, then $\mathit{VA}$ is resized
together with the user variable section. In this case the selector variables
are assigned new, larger IDs and copied to a new position in
$\mathit{VA}$. Then the literals of selector variables are renamed according
to the newly assigned IDs in all (learned) clauses and cubes present in the
current PCNF in a single pass.  Resizing only the selector variable section of
$\mathit{VA}$ does not require assigning new IDs to selector
variables. Similar to implementations of other SAT or QBF solvers, the user is
responsible to avoid unnecessarily large user variable indices and thus avoid
resizing $\mathit{VA}$.

The API of DepQBF prevents accessing selector variables in $\mathit{VA}$,
which are hence invisible to the user. In contrast to traditional solver
implementations, e.g.~\cite{DBLP:journals/entcs/EenS03}, where the user is
responsible to maintain selector variables manually, the \emph{internal}
separation between user and selector variables allows to conveniently allocate
and rename selector variables on the fly inside the solver and without any
user interaction. This feature is particularly useful for solving dynamically generated
sequences $S = \langle \psi_1,\ldots,\psi_n\rangle$ of PCNFs where the exact
user variable IDs in each $\psi_i$ are unknown at the
beginning.

In the following, we present the novel clause group API of DepQBF along with
the example shown in Fig.~\ref{fig_code}.
\begin{figure}[t]
\begin{minipage}{0.52\textwidth}
{
\footnotesize
\begin{verbatim}
int main (int argc, char ** argv) {
  Solver *s = create();
  new_scope_at_nesting 
    (s,QTYPE_FORALL,1);
  add(s,1);add(s,2);add(s,0);
  new_scope_at_nesting 
    (s,QTYPE_EXISTS,2);
  add(s,3);add(s,4);add(s,0);

  ClauseGroupID id1 = new_cls_grp(s);
  open_cls_grp(s,id1);
  add(s,-1);add(s,-3);add(s,0);
  close_cls_grp(s,id1);

  ClauseGroupID id2 = new_cls_grp(s);
  open_cls_grp(s,id2);
  add(s,1);add(s,2);add(s,4);add(s,0);
  add(s,1);add(s,-4);add(s,0);
  close_cls_grp(s,id2);
  ...//continues on right column.
\end{verbatim}
}
\end{minipage}
\begin{minipage}{0.47\textwidth}
{
\footnotesize
\begin{verbatim}
  ...//continued from left column.
  Result res = sat(s);
  assert(res == RESULT_UNSAT);
  ClauseGroupID *rgrps = 
    get_relevant_cls_grps(s);
  assert(rgrps[0] == id2);
  reset(s);

  deactivate_cls_grp(s,rgrps[0]);
  res = sat(s);
  assert(res == RESULT_SAT);
  reset(s);

  activate_cls_grp(s,rgrps[0]);
  free(rgrps);

  delete_cls_grp(s,id1);
  res = sat(s);
  assert(res == RESULT_UNSAT);
  delete(s); }
\end{verbatim}
}
\end{minipage}
\caption{Clause group code example. Variables $x_i$ are encoded as integers
$i$.  Given the PCNF $\psi := \forall x_1,x_2\exists x_3,x_4.\,C_1 \wedge C_2
\wedge C_3$ with $C_1 = (\neg x_1 \vee \neg x_3), C_2 = (x_1 \vee x_2 \vee
x_4), C_3 = (x_1 \vee \neg x_4)$, $C_1$ is put in group \texttt{id1} and
$C_2,C_3$ in group \texttt{id2}. An unsatisfiable core consisting only of
group \texttt{id2} is extracted from $\psi$. Deactivating group \texttt{id2}
results in the PCNF $\forall x_1\exists x_3.\,C_1$. Activating \texttt{id2}
again and deleting \texttt{id1} yields $\forall x_1,x_2\exists x_4.\,C_2
\wedge C_3$.}
\label{fig_code}
\end{figure}
A new clause group is created by calling \texttt{new\_cls\_grp()}, which
returns a unique unsigned integer $\mathit{cgid}$ as the ID of the group. Each
time a new group $\mathit{cgid}$ is created, \emph{internally} a fresh
selector variable $s$ is allocated in the array $\mathit{VA}$ and associated
with the group $\mathit{cgid}$.

A group $\mathit{cgid}$ must be opened by \texttt{open\_cls\_grp(cgid)} before
clauses can be added to it. All clauses added via the API are associated with
the currently opened group $\mathit{cgid}$ by \emph{internally} augmenting
them with the selector variable $s$ of group $\mathit{cgid}$.  Groups must be
closed by \texttt{close\_cls\_grp(cgid)}.  When solving the current PCNF by
\texttt{sat()}, \emph{internally} the selector variables of all created groups
are assigned \emph{false} as assumptions to effectively activate the clauses
in these groups.

Deleting a group by \texttt{delete\_cls\_grp(cgid)} invalidates its ID. When
solving the current PCNF by \texttt{sat()}, \emph{internally} the selector
variables of all deleted groups are assigned \emph{true} as assumptions to
deactivate the clauses in all deleted groups and all learned clauses derived
therefrom. Deleted clauses are physically removed from the data structures in
a garbage collection phase if their number exceeds a certain threshold.
Clauses which are added to the PCNF without opening a group by
\texttt{open\_cls\_grp(cgid)} before are permanent and cannot be deleted.

In contrast to deletion, clause groups can also be \emph{deactivated} by
calling \texttt{deactivate\_cls\_grp(cgid)}.  When solving the current PCNF by
\texttt{sat()}, \emph{internally} the selector variables of deactivated groups
are assigned \emph{true} similarly to deleted groups. However, clauses in
deactivated groups are never removed from the data structures. Deactivated
groups are activated again by \texttt{activate\_cls\_grp(cgid)}.  Selector
variables of activated groups are assigned \emph{false} when solving the
current PCNF.

DepQBF also allows for traditional incremental solving where the user handles
selector variables manually~\cite{DBLP:journals/entcs/EenS03}. Implementations
of this approach like MiniSAT, for example, allow to physically delete clauses
by first adding a unit clause containing a selector variable and then
simplifying the formula based on unit clauses. This is in contrast to DepQBF
where the formula is not simplified based on unit clauses to avoid the internal
elimination of variables, which may be unexpected by the user.

If the current PCNF has been found unsatisfiable by \texttt{sat()}, then
calling \texttt{get\_relevant\_cls\_grps()} returns an array of the IDs of
those groups which contain clauses used by the solver to determine
unsatisfiability. The clauses in these groups amount to an unsatisfiable core
of the PCNF. That core is obtained by \emph{internally} collecting all
selector variables relevant for unsatisfiability\footnote{Similar to the
function \texttt{analyzeFinal} in Minisat, for example.} and mapping them to
the respective clause group IDs.


\section{Computing Minimal Unsatisfiable Cores of QBFs} \label{sec_experiments}

In contrast to theory~\cite{DBLP:conf/sat/BuningZ06}, the computation of MUCs
of PCNFs in \emph{practice} has not been considered so far. Approaches to
\emph{nonminimal} UCs of PCNFs were presented in the context of checking
Q-resolution refutations of PCNFs~\cite{DBLP:conf/aspdac/YuM05} and
QMaxSAT~\cite{DBLP:conf/sat/IgnatievJM13}.  For the first time we report on
experiments related to the computation of MUCs of PCNFs.  To this end, we
implemented a tool to incrementally compute MUCs of PCNFs using the clause
group API of DepQBF as follows.

Given an unsatisfiable PCNF $\psi_0 = \hat{Q}.\,\phi$, first every single
clause of $\psi_0$ is put in an individual clause group. Let $\psi :=
\psi_0$. The PCNF $\psi$ is solved and a UC $\psi' = \hat{Q}'.\,\phi'$ is
extracted by \texttt{get\_relevant\_cls\_grps}. Then $\psi$ is replaced by
$\psi'$ by deleting the clause groups which do not belong to $\psi'$ from
$\psi$. Given the updated $\psi = \hat{Q}.\,\phi$, every clause $C \in \phi$
is checked by solving the PCNF $\psi'' = \hat{Q}.\,(\phi \setminus \{C\})$.
To this end, the group containing $C$ is deactivated. If $\psi''$ is
satisfiable then $C$ is part of an MUC and hence $C$ is activated again ($C$
is a \emph{transition clause}~\cite{DBLP:conf/sat/SilvaL11}).  Otherwise, a UC
$\psi'$ of $\psi''$ is extracted, $\psi$ is replaced by the UC $\psi'$ like
above, and again every clause in the updated $\psi$ is checked. After every
clause in the current $\psi$ has been checked, the final $\psi$ is an MUC of
$\psi_0$. The number of solver calls in this well-known elimination-based
algorithm is linear in the size of
$\psi_0$~\cite{DBLP:conf/ictai/GregoireMP08,DBLP:conf/ismvl/Silva10,DBLP:conf/sat/SilvaL11}. It
applies iterative \emph{clause set
refinement}~\cite{DBLP:journals/aicom/BelovLM12,DBLP:conf/sat/DershowitzHN06,DBLP:conf/fmcad/Nadel10}
by UCs. UCs are extracted by selector
variables~\cite{DBLP:conf/lpar/AsinNOR08} in
\texttt{get\_relevant\_cls\_grps}, which is in contrast to extraction based on
resolution
proofs~\cite{DBLP:conf/fmcad/Nadel10,DBLP:journals/jsat/NadelRS14}. The
algorithm is common to compute MUCs of CNFs but has not been applied to PCNFs
so far.

Using our tool, we computed MUCs of instances from the \emph{applications
(AT)}, \emph{QBFLIB (QT)}, and \emph{preprocessing (PT)} tracks of the QBF
Gallery 2014.\footnote{\url{http://qbf.satisfiability.org/gallery/}} We 
preprocessed the instances from \emph{AT} and \emph{QT} using 
Bloqqer~\cite{DBLP:conf/cade/BiereLS11}. In total, we allowed 900s of wall
clock time and seven GB of memory to solve an instance by DepQBF and to
compute an MUC. Table~\ref{tab_mus} summarizes the results of our
experiments run on an AMD
Opteron 6238 at 2.6~GHz under 64-bit Linux.
\begin{table}[t]
\centering
\begin{tabular}{lr@{\hspace{0.2cm}}r@{\hspace{0.2cm}}r@{\hspace{0.2cm}}r@{\hspace{0.2cm}}r@{\hspace{0.2cm}}rrr}
\hline
\multicolumn{1}{c}{\emph{QBF Gallery Track}} & \multicolumn{1}{c@{\hspace{0.2cm}}}{\emph{\#m}} & \multicolumn{1}{c@{\hspace{0.2cm}}}{\emph{ut}} & \multicolumn{1}{c@{\hspace{0.2cm}}}{\emph{mt}} &
\multicolumn{1}{c@{\hspace{0.2cm}}}{$|\mathit{CNF}|$} & \multicolumn{1}{c@{\hspace{0.2cm}}}{$|\mathit{MUC}|$} & \multicolumn{1}{c}{\emph{\#c}} & \multicolumn{1}{c}{$\overline{r}$} & \multicolumn{1}{c}{$\tilde{r}$}\\
\hline
\emph{applications (190 of 735):} & 182 & 6,304 & 7,941 & 4,744,494  & 73,206 & 81,631 & 6.1\% & 2.9\% \\
\emph{QBFLIB \hspace{0.475cm} (58 of 276):}  & 46 & 1,009 & 2,264 & 323,497 & 34,777 & 36,888 & 14.1\% & 5.1\% \\
\emph{preprocessing (38 of 243):}  & 34 & 1,623 & 1,080 & 451,197 & 23,220 & 24,572 & 4.0\% & 2.2\% \\
\hline
\end{tabular}
\caption{Statistics for unsatisfiable instances from the QBF Gallery 2014 where MUCs were
successfully computed. Numbers of solved instances out of total
ones are shown in parentheses. MUCs computed (\emph{\#m}), total time to solve
the initial unsatisfiable instances (\emph{ut}) and to compute the MUCs
(\emph{mt}), total number of
clauses in initial formulas ($|\mathit{CNF}|$) and in MUCs ($|\mathit{MUC}|$), 
total number of QBF solver calls (\emph{\#c}), 
and the average ($\overline{r}$) and median ($\tilde{r}$) sizes of MUCs
relative to the respective CNF sizes.}
\label{tab_mus}
\end{table}
MUCs were successfully computed for 95\% of the solved unsatisfiable instances
in \emph{AT} (79\% of \emph{QT} and 89\% of \emph{PT}). On average, MUC
computation took 43s in \emph{AT} (49s in \emph{QT} and 31s in
\emph{PT}). When increasing the total timeout to 3600s, then 186 MUCs were
computed in \emph{AT} (48 in \emph{QT} and 36 in \emph{PT}).

Iterative clause set refinement by UCs potentially reduces the number of
solver calls. In the worst case, there is one solver call per each single
clause in the initial PCNF $\psi_0$. However, on average there was one solver
call per $58$, $8$, and $18$ clauses in \emph{AT}, \emph{QT}, and \emph{PT},
respectively.

The physical deletion of clauses not belonging to a MUC reduces the memory
footprint and the run time. The plot below shows the sorted total run times
(y-axis) of the MUC workflow on instances in \emph{AT} where MUCs were
successfully computed (x-axis). If clauses are deleted by
\texttt{delete\_cls\_grp} (UC-d) then 182 MUCs are computed but only 169 if
clauses are permanently deactivated by \texttt{deactivate\_cls\_grp} instead
(UC-nd). We attribute this effect to
\begin{wrapfigure}{r}{0.445\textwidth}
\vspace{-0.67cm}
\includegraphics[scale=0.5]{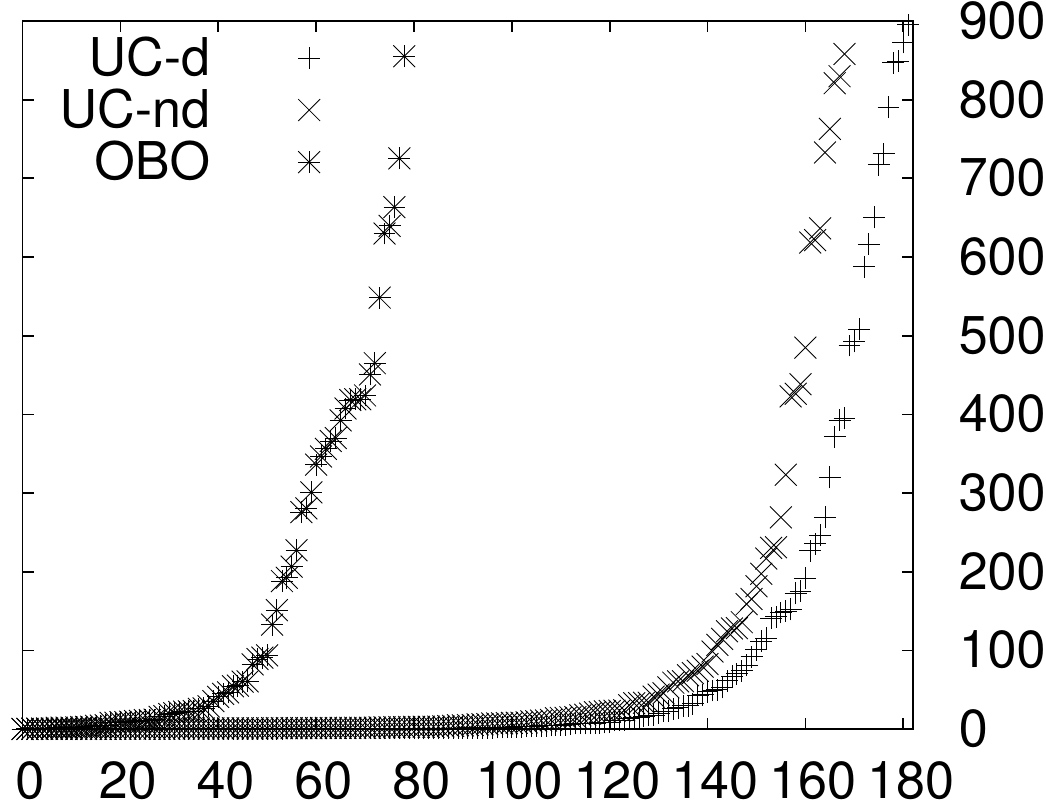}
\vspace{-0.5cm}
\end{wrapfigure}
overhead caused by deactivated clauses still present in the data
structures. Only 79 MUCs are computed without iterative clause set refinement
by UCs using \texttt{get\_relevant\_cls\_grps} and instead checking
\emph{every} clause in $\psi_0$ one by one (OBO). We made similar observations
for \emph{QT} and \emph{PT}.  On instances where an MUC was computed by both
UC-d and UC-nd, in general UC-nd is slower (up to $+316\%$ on \emph{PT}) and
has a larger memory footprint (up to $+70\%$ on \emph{AT}). The difference
between UC-d and OBO is more pronounced, where in general OBO is slower (up to
$+4126\%$ on \emph{PT}) and has a larger memory footprint (up to $+243\%$ on
\emph{AT}).

Our experiments show that physical deletion of clauses by
\texttt{delete\_cls\_grp} (UC-d) and the extraction of UCs by
\texttt{get\_relevant\_cls\_grps} based on selector variables are crucial for
the computation of MUCs of PCNFs. These features are provided directly by the
novel clause group API of DepQBF.


\section{Conclusion}

We presented a novel API of our solver DepQBF for incremental QBF solving
based on clause groups and its application to MUC computation. The clause
group API is conceptually simple yet employs state of the art approaches to
assumption-based incremental SAT solving. Improvements of
assumption-based incremental
solving~\cite{DBLP:conf/sat/AudemardLS13,DBLP:conf/sat/LagniezB13,DBLP:conf/sat/NadelRS14}
are also applicable to our implementation.

The API encapsulates the handling of selector variables and assumptions
entirely inside the solver. This is a unique feature of DepQBF, which
facilitates its integration in other tools. It is particularly useful for
solving dynamically generated sequences of PCNFs where the exact variable IDs are unknown
at the beginning.  The clause group API is general and fits any search-based
SAT and QBF solver capable of solving under assumptions.

A potential application of the clause group API is (M)UC extraction of PCNFs
in core-guided QMaxSAT~\cite{DBLP:conf/sat/IgnatievJM13} and SMT, similar to
SAT-based UC extraction in SMT~\cite{DBLP:conf/sat/CimattiGS07}. Further, our
API readily supports the extraction of high-level
UCs~\cite{DBLP:journals/jar/LiffitonS08,DBLP:conf/fmcad/Nadel10,DBLP:journals/jsat/NadelRS14}
where, different from our experiments with MUC computation, multiple clauses
are put in a clause group.  We applied the novel clause group API of DepQBF to
compute MUCs of PCNFs for the first time. Our results indicate the efficiency
and applicability of our implementation. As future work, we want to integrate
incremental preprocessing in DepQBF in a way where the implementation details
are hidden by the API~\cite{DBLP:conf/sat/NadelRS14}.



\newpage

\begin{appendix}

\section{Additional Experimental Data}

\begin{table}
\centering
\begin{tabular}{lr@{\hspace{0.2cm}}r@{\hspace{0.2cm}}r@{\hspace{0.2cm}}r@{\hspace{0.2cm}}r@{\hspace{0.2cm}}rrr}
\hline
\multicolumn{1}{c}{\emph{QBF Gallery Track}} & \multicolumn{1}{c@{\hspace{0.2cm}}}{\emph{\#m}} & \multicolumn{1}{c@{\hspace{0.2cm}}}{\emph{ut}} & \multicolumn{1}{c@{\hspace{0.2cm}}}{\emph{mt}} &
\multicolumn{1}{c@{\hspace{0.2cm}}}{$|\mathit{CNF}|$} & \multicolumn{1}{c@{\hspace{0.2cm}}}{$|\mathit{MUC}|$} & \multicolumn{1}{c}{\emph{\#c}} & \multicolumn{1}{c}{$\overline{r}$} & \multicolumn{1}{c}{$\tilde{r}$}\\
\hline
\emph{applications (190 of 735):} & 169 & 4448 & 7346 & 1,604,908 & 46,025 & 53,080 & 6.4\% & 3.0\% \\
\emph{QBFLIB \hspace{0.475cm} (58 of 276):} & 46 & 982 & 3,308 & 323,497 & 34,832 & 37,020 & 14.1\% & 5.2\% \\
\emph{preprocessing (38 of 243):} & 31 & 613 & 1658 & 368,002 & 14,673 & 15,913 & 3.4\% & 2.2\% \\
\hline
\end{tabular}
\caption{Related to Table~\ref{tab_mus} (same column labels): computation of
MUCs where groups of clauses not being part of a MUC are deactivated instead
of deleted (``UC-nd'' in the plot in Section~\ref{sec_experiments}). Except
for the \emph{QBFLIB} track, fewer MUCs were computed when deactivating
clauses than when deleting clauses.}
\label{tab_mus_no_delete}
\end{table}

\begin{table}
\centering
\begin{tabular}{lr@{\hspace{0.2cm}}r@{\hspace{0.2cm}}r@{\hspace{0.2cm}}r@{\hspace{0.2cm}}r@{\hspace{0.2cm}}rrr}
\hline
\multicolumn{1}{c}{\emph{QBF Gallery Track}} & \multicolumn{1}{c@{\hspace{0.2cm}}}{\emph{\#m}} & \multicolumn{1}{c@{\hspace{0.2cm}}}{\emph{ut}} & \multicolumn{1}{c@{\hspace{0.2cm}}}{\emph{mt}} &
\multicolumn{1}{c@{\hspace{0.2cm}}}{$|\mathit{CNF}|$} & \multicolumn{1}{c@{\hspace{0.2cm}}}{$|\mathit{MUC}|$} & \multicolumn{1}{c}{\emph{\#c}} & \multicolumn{1}{c}{$\overline{r}$} & \multicolumn{1}{c}{$\tilde{r}$}\\
\hline
\emph{applications (190 of 735):} & 79 & 238 & 12,004 & 310,986 & 22,367 & 311,065 & 10.6\% & 3.8\% \\
\emph{QBFLIB \hspace{0.475cm} (58 of 276):}  & 34 & 608 & 5,970 & 169,853 & 20,463 & 169,887 & 17.4\% & 7.9\% \\
\emph{preprocessing (38 of 243):}  & 21 & 270 & 5,326 & 175,162 & 5,233 & 175,183 & 3.2\% & 2.6\% \\
\hline
\end{tabular}
\caption{Related to Table~\ref{tab_mus} (same column labels): computation of
MUCs without iterative clause set refinement by UCs but instead checking every
clause in the initial PCNF one by one (``OBO'' in the plot in
Section~\ref{sec_experiments}). This approach performs considerably worse than
either variant of iterative clause set refinement by UCs (UC-d and UC-nd in
Tables~\ref{tab_mus} and~\ref{tab_mus_no_delete}).}
\label{tab_mus_one_by_one}
\end{table}

\begin{table}
\centering
\begin{tabular}{lr@{\hspace{0.2cm}}r@{\hspace{0.2cm}}r@{\hspace{0.2cm}}r@{\hspace{0.2cm}}r@{\hspace{0.2cm}}r@{\hspace{0.2cm}}r@{\hspace{0.2cm}}r@{\hspace{0.2cm}}r@{\hspace{0.2cm}}}
\hline
\multicolumn{1}{c}{\emph{}} & \multicolumn{1}{c}{\emph{\#m}} &
\multicolumn{1}{c}{\emph{mt}} & \multicolumn{1}{c}{\emph{mem}} &
\multicolumn{1}{c}{$|\mathit{CNF}|$} & \multicolumn{1}{c}{$|\mathit{MUC}|$} &
\multicolumn{1}{c}{\emph{\#c}} & \multicolumn{1}{c}{$\overline{r}$} &
\multicolumn{1}{c}{$\tilde{r}$} \\
\hline
\emph{UC-d:}      & 46  & 2,264 & 2,175 & 323,497 & 34,777 & 36,888 & 14.1\% & 5.1\% \\
\emph{UC-nd:}  & 46  & 3,308 & 2,948 & 323,497 & 34,832 & 37,020 & 14.1\% & 5.2\% \\
\hline
\end{tabular}
\caption{Related to Table~\ref{tab_mus} (same column labels except \emph{mem}
which is the total amount of memory) and to the plot in
Section~\ref{sec_experiments}: comparison of the MUC workflow on instances
from the \emph{QBFLIB} track where a MUC was computed by both iterative clause
set refinement by UCs where clauses are deleted (\emph{UC-d}) and deactivated
(\emph{UC-nd}), respectively. Whereas the sizes of MUCs and numbers of solver
calls are similar for \emph{UC-d} and \emph{UC-nd}, \emph{UC-nd} is slower
($+46\%$) and has a larger memory footprint ($+35\%$).}
\label{tab_mus_intersection_qbflib}
\end{table}

\begin{table}
\centering
\begin{tabular}{lr@{\hspace{0.2cm}}r@{\hspace{0.2cm}}r@{\hspace{0.2cm}}r@{\hspace{0.2cm}}r@{\hspace{0.2cm}}r@{\hspace{0.2cm}}r@{\hspace{0.2cm}}r@{\hspace{0.2cm}}r@{\hspace{0.2cm}}}
\hline
\multicolumn{1}{c}{\emph{}} & \multicolumn{1}{c}{\emph{\#m}} &
\multicolumn{1}{c}{\emph{mt}} & \multicolumn{1}{c}{\emph{mem}} &
\multicolumn{1}{c}{$|\mathit{CNF}|$} & \multicolumn{1}{c}{$|\mathit{MUC}|$} &
\multicolumn{1}{c}{\emph{\#c}} & \multicolumn{1}{c}{$\overline{r}$} &
\multicolumn{1}{c}{$\tilde{r}$} \\
\hline
\emph{UC-d:}   & 31  & 398 & 1,672 & 368,002 & 14,586 & 15,785 & 3.4\% & 2.2\% \\
\emph{UC-nd:}  & 31  & 1,658 & 1,988 & 368,002 & 14,673 & 15,913 & 3.4\% & 2.2\% \\
\hline
\end{tabular}
\caption{Like Table~\ref{tab_mus_intersection_qbflib} but for instances from
the \emph{preprocessing} track. \emph{UC-nd} is slower ($+316\%$) and has a
larger memory footprint ($+18\%$).}
\label{tab_mus_intersection_prepro}
\end{table}

\begin{table}
\centering
\begin{tabular}{lr@{\hspace{0.2cm}}r@{\hspace{0.2cm}}r@{\hspace{0.2cm}}r@{\hspace{0.2cm}}r@{\hspace{0.2cm}}r@{\hspace{0.2cm}}r@{\hspace{0.2cm}}r@{\hspace{0.2cm}}r@{\hspace{0.2cm}}}
\hline
\multicolumn{1}{c}{\emph{}} & \multicolumn{1}{c}{\emph{\#m}} &
\multicolumn{1}{c}{\emph{mt}} & \multicolumn{1}{c}{\emph{mem}} &
\multicolumn{1}{c}{$|\mathit{CNF}|$} & \multicolumn{1}{c}{$|\mathit{MUC}|$} &
\multicolumn{1}{c}{\emph{\#c}} & \multicolumn{1}{c}{$\overline{r}$} &
\multicolumn{1}{c}{$\tilde{r}$} \\
\hline
\emph{UC-d:}      & 169 & 4,118 & 5,155 & 1,604,908 & 44,979 & 52,049 & 6.4\% & 2.9\% \\
\emph{UC-nd:}  & 169 & 7,346 & 8,779 & 1,604,908 & 46,025 & 53,080 & 6.4\% & 3.0\% \\
\hline
\end{tabular}
\caption{Like Table~\ref{tab_mus_intersection_qbflib} but for instances from
the \emph{applications} track. \emph{UC-nd} is slower ($+78\%$) and has a
larger memory footprint ($+70\%$).}
\label{tab_mus_intersection_applications}
\end{table}

\begin{table}
\centering
\begin{tabular}{lr@{\hspace{0.2cm}}r@{\hspace{0.2cm}}r@{\hspace{0.2cm}}r@{\hspace{0.2cm}}r@{\hspace{0.2cm}}r@{\hspace{0.2cm}}r@{\hspace{0.2cm}}r@{\hspace{0.2cm}}r@{\hspace{0.2cm}}}
\hline
\multicolumn{1}{c}{\emph{}} & \multicolumn{1}{c}{\emph{\#m}} &
\multicolumn{1}{c}{\emph{mt}} & \multicolumn{1}{c}{\emph{mem}} &
\multicolumn{1}{c}{$|\mathit{CNF}|$} & \multicolumn{1}{c}{$|\mathit{MUC}|$} &
\multicolumn{1}{c}{\emph{\#c}} & \multicolumn{1}{c}{$\overline{r}$} &
\multicolumn{1}{c}{$\tilde{r}$} \\
\hline
\emph{UC-d:} & 34 & 1,235 & 1,040 & 169,853 & 19,603 & 21,192 & 15.7\% & 5.1\% \\
\emph{OBO:}  & 34 & 5,970 & 1,743 & 169,853 & 20,463 & 169,887 & 17.4\% & 7.9\% \\
\hline
\end{tabular}
\caption{Related to Table~\ref{tab_mus} (same column labels except \emph{mem}
which is the total amount of memory) and to the plot in
Section~\ref{sec_experiments}: comparison of the MUC workflow on instances
from the \emph{QBFLIB} track where a MUC was computed by both iterative clause
set refinement by UCs where clauses are deleted (\emph{UC-d}) and without
iterative clause set refinement and checking clauses one by one instead
(\emph{OBO}), respectively. In addition to the solver calls for solving the
initial unsatisfiable PCNFs, with \emph{OBO} there is one solver call for each
clause in the initial PCNF. \emph{OBO} is slower ($+383\%$) and has a larger
memory footprint ($+67\%$).}
\label{tab_mus_intersection_qbflib_relevant_one_by_one}
\end{table}

\begin{table}
\centering
\begin{tabular}{lr@{\hspace{0.2cm}}r@{\hspace{0.2cm}}r@{\hspace{0.2cm}}r@{\hspace{0.2cm}}r@{\hspace{0.2cm}}r@{\hspace{0.2cm}}r@{\hspace{0.2cm}}r@{\hspace{0.2cm}}r@{\hspace{0.2cm}}}
\hline
\multicolumn{1}{c}{\emph{}} & \multicolumn{1}{c}{\emph{\#m}} &
\multicolumn{1}{c}{\emph{mt}} & \multicolumn{1}{c}{\emph{mem}} &
\multicolumn{1}{c}{$|\mathit{CNF}|$} & \multicolumn{1}{c}{$|\mathit{MUC}|$} &
\multicolumn{1}{c}{\emph{\#c}} & \multicolumn{1}{c}{$\overline{r}$} &
\multicolumn{1}{c}{$\tilde{r}$} \\
\hline
\emph{UC-d:} & 21 & 126 & 850 & 175,162 & 4,977 & 5,861 & 3.2\% & 2.2\% \\
\emph{OBO:}  & 21 & 5,326 & 1,135 & 175,162 & 5,233 & 175,183 & 3.2\% & 2.6\% \\
\hline
\end{tabular}
\caption{Like Table~\ref{tab_mus_intersection_qbflib_relevant_one_by_one} but
for instances from the \emph{preprocessing} track. \emph{OBO} is slower
($+4126\%$) and has a larger memory footprint ($+33\%$).}
\label{tab_mus_intersection_prepro_relevant_one_by_one}
\end{table}

\begin{table}
\centering
\begin{tabular}{lr@{\hspace{0.2cm}}r@{\hspace{0.2cm}}r@{\hspace{0.2cm}}r@{\hspace{0.2cm}}r@{\hspace{0.2cm}}r@{\hspace{0.2cm}}r@{\hspace{0.2cm}}r@{\hspace{0.2cm}}r@{\hspace{0.2cm}}}
\hline
\multicolumn{1}{c}{\emph{}} & \multicolumn{1}{c}{\emph{\#m}} &
\multicolumn{1}{c}{\emph{mt}} & \multicolumn{1}{c}{\emph{mem}} &
\multicolumn{1}{c}{$|\mathit{CNF}|$} & \multicolumn{1}{c}{$|\mathit{MUC}|$} &
\multicolumn{1}{c}{\emph{\#c}} & \multicolumn{1}{c}{$\overline{r}$} &
\multicolumn{1}{c}{$\tilde{r}$} \\
\hline
\emph{UC-d:} & 79 & 1,724 & 839 & 310,986 & 19,878 & 23,665 & 10.3\% & 2.9\% \\
\emph{OBO:} & 79 & 12,004 & 2,879 & 310,986 & 22,367 & 311,065 & 10.6\% & 3.8\% \\
\hline
\end{tabular}
\caption{Like Table~\ref{tab_mus_intersection_qbflib_relevant_one_by_one} but
for instances from the \emph{applications} track. \emph{OBO} is slower
($+596\%$) and has a larger memory footprint ($+243\%$).}
\label{tab_mus_intersection_applications_relevant_one_by_one}
\end{table}

\begin{figure}
\centering
\includegraphics[scale=0.5]{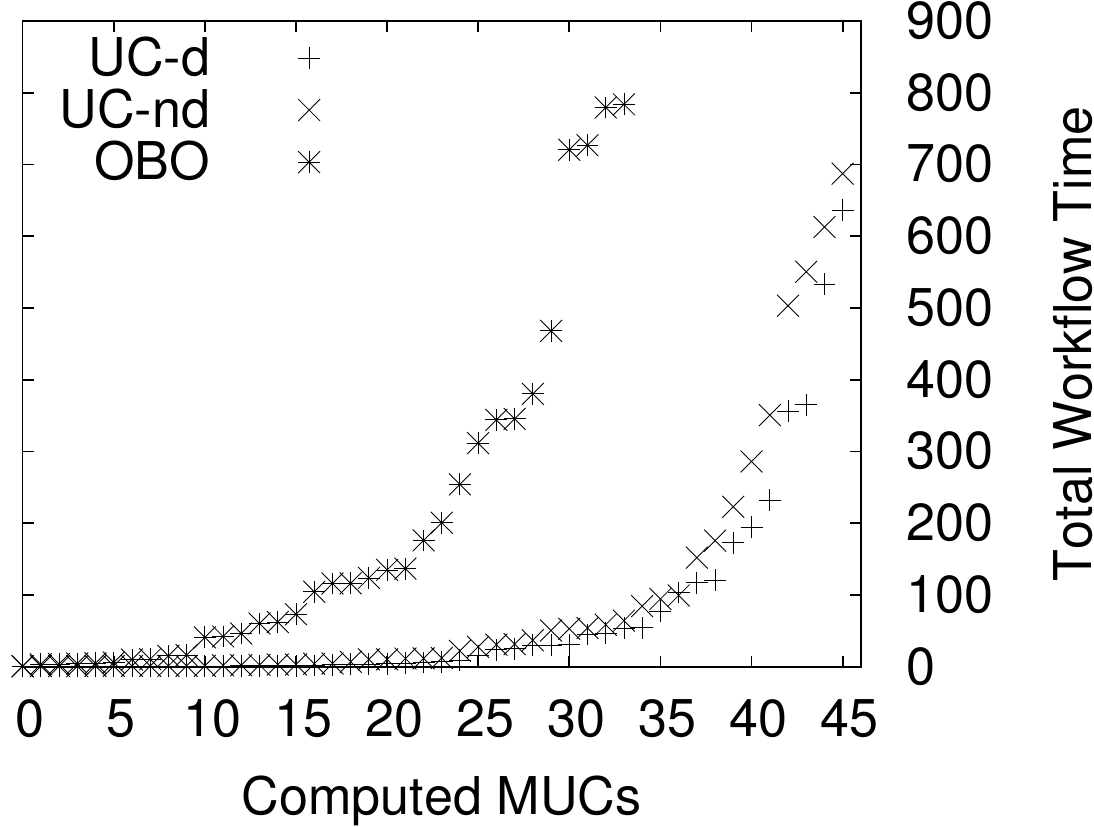} 
\caption{Like the plot in
  Section~\ref{sec_experiments} but for instances from the \emph{QBFLIB} track.}
\label{plot_qbflib}
\end{figure}
\begin{figure}
\centering
\includegraphics[scale=0.5]{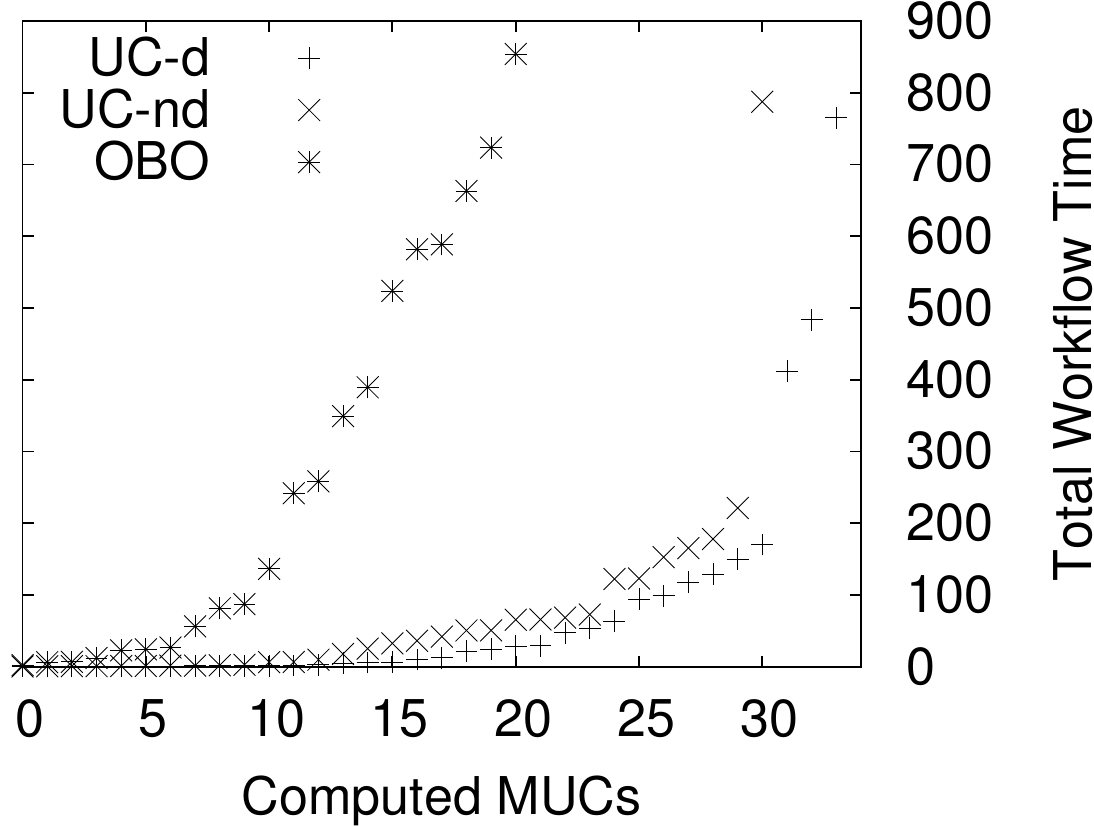}
\caption{Like the plot in Section~\ref{sec_experiments} but for instances from
the \emph{preprocessing} track.}
\label{plot_prepro}
\end{figure}

\end{appendix}


\end{document}